\documentstyle[twoside,fleqn,epsfig,espcrc2]{article}

\newcommand{\AmS}{{\protect\the\textfont2
  A\kern-.1667em\lower.5ex\hbox{M}\kern-.125emS}}

\title{Simulation of Water Cerenkov Detectors Using {\sc geant4}}

\author{L. A. Anchordoqui\address{Department of Physics, 
        Northeastern University, Boston, 
        MA 02115, USA}\thanks{doqui@hepmail.physics.neu.edu}, 
        T. P. McCauley$^a$\thanks{mccauley@hepmail.physics.neu.edu},   
        T. Paul$^a$\thanks{tom.paul@hepmail.physics.neu.edu},
        S. Reucroft$^a$\thanks{stephen.reucroft@cern.ch},  
        J. D. Swain$^a$\thanks{john.swain@cern.ch}, $\,$ and 
        L. Taylor$^a$\thanks{lucas.taylor@cern.ch}}

\begin{document}

\begin{abstract}

We present a detailed simulation of the performance
of water Cerenkov detectors suitable for use in the Pierre Auger
Observatory. Using {\sc geant4}, a flexible object-oriented simulation
program, including all known physics processes, has been developed.
The program also allows interactive visualization, and can easily
be modified for any experimental setup.
\end{abstract}

\maketitle

%
%
%
%
%
%
%

\section{Introduction}
Water Cerenkov detectors have proved to be superb devices for the study of cosmic 
air showers~\cite{other_experiments}, and will constitute major components
of future experiments such as the Pierre Auger Observatory~\cite{auger}.  Monte 
Carlo simulation of the response of such detectors to signal and background processes
is crucial for proper interpretation of the data and to aid in development of
reconstruction and analysis algorithms. To address these needs, experiments
have prepared dedicated detector simulations~\cite{agasim}.  There 
may be advantages, however, in exploiting some of the efforts which the High Energy
Physics (HEP) community has directed at this problem.  The {\sc geant3} package~\cite{geant3}, 
for example, was developed in order to provide HEP experiments with generic tools 
for simulating the passage of particles through matter, but it also found substantial 
usage in the medical and biological sciences and in astronautics, and
in fact has been studied for use in simulating water Cerenkov detectors for the Auger
experiment~\cite{g3auger}.  In 1996 CERN initiated 
the {\sc geant4} project~\cite{geant4} with the goal of reproducing all the functionality of 
{\sc geant3} using an Object Oriented approach, as well as addressing some of the shortfalls
of the older program.  We have begun investigating the suitability of {\sc geant4} for the
problem of simulating water Cerenkov detectors, and report here on the status of the
work. 

\section{{\sc geant4}}

Like its predecessor, {\sc geant4} provides a battery of tools to describe the 
geometry and material properties of an experimental setup, handle particle 
transport through materials and magnetic fields, and simulate particle decay 
and interactions with detector elements.  All relevant physics processes 
have been included. For example, simulation of an electron traversing a 
material can include effects of ionisation, delta ray production,
multiple coulomb scattering, bremsstrahlung and Cerenkov radiation.  
Optical photons produced by processes such as Cerenkov radiation may then
be subjected to Rayleigh scattering, absorption, and optical boundary 
interactions.  We note that the calculation of reflection and transmission
coefficients at material boundaries takes into account the polarization 
state of the photon, which is important in accounting for a photon's fate
as it traverses multiple optical boundaries en route to a photocathode.  
 
In addition to reproducing the functionality of {\sc geant3}, {\sc geant4} aims to 
improve the procedures used for geometry definition, introduction of special physics 
processes, visualization, and optical processes. In addition, special 
effort has been made to ensure tracking precision over arbitrary scales,
so the package may lend itself to simulation of extensive air showers~\cite{geantshowers}
as well as detailed detector response to ground particles.
{\sc geant4} is written in C++ using an object oriented approach, and as such 
alleviates some of the shortcomings inherent in the procedural approach (FORTRAN77) 
of {\sc geant3}.

\section{Detector Geometry}
\begin{figure}
\begin{center}
\epsfig{file=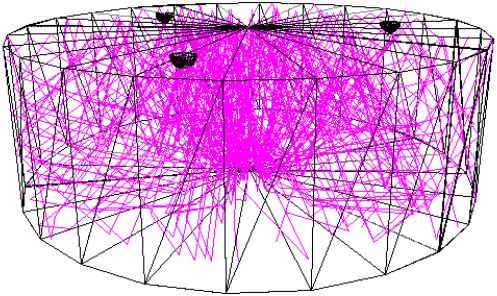,width=7cm,clip=}
\caption{\label{fig:tank}A muon entering a tank of water from above and radiating 
Cerenkov light. Housings
for three photomultiplier tubes are visible at the top of the tank.}
\end{center}
\end{figure}

{\sc geant4} strives to provide rather advanced tools for describing 
detector geometry, such as methods for interpreting files produced by 
Computer Aided Design (CAD) systems.
However, for our application a first approximation to the geometry may be
described in only a few lines of code.  
Figure~\ref{fig:tank} shows an example of an Auger-style Cerenkov detector
which consists of a cylindrical tank filled with water.
A reflective liner surrounds the water volume.  Three 
hemispherical domes at the top of the tank house photomultiplier 
tubes, which contain sensitive photocathode volumes that register a hit when
a photon strikes.  Photons bounce about inside the tank until they are absorbed
in the water, the tank walls, or until they enter a photomultiplier.  The 
frequency-dependent optical properties of all these elements can be tuned
to available data. In the figure, a 1~GeV muon is incident at the top of the 
tank, and some of the radiated Cerenkov photons are shown.

\section{Optical Modeling}
\begin{figure}
\begin{center}
\epsfig{file=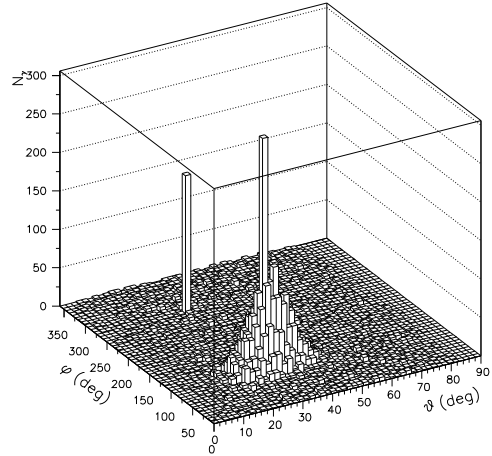,width=7cm,clip=}
\caption{\label{fig:light} Distribution of reflected photons as a function of polar angles 
$\theta$ and $\phi$ for a beam incident at $\theta_{\mathrm incident} = 30^\circ$. 
This simulation uses the unified model of
{\sc geant4} with specular and backscatter spikes, visible at $\phi = 90^\circ$ and $270^\circ$ respectively,
a specular lobe centered at $\theta = 30^\circ$, as well as uniform (Lambertian) reflection.}
\end{center}
\end{figure}
The detector geometry of Figure~\ref{fig:tank} includes a diffuse reflector to 
contain the Cerenkov light inside the volume of water.  The use of 
Tyvek~\footnote{Tyvek is a trademark of DuPont.} as such a reflector 
has been explored by the Auger experiment, and it has been 
noted~\cite{tyvek1} that the treatment of photon interactions with a rough 
dielectric surface as available in {\sc geant3} does not appear to be sufficient to 
describe the experimentally observed reflection of light from this material.
{\sc geant4} provides a more flexible optical model inspired by the work of 
Nayer {\it et al.}~\cite{nayer}. There it was observed that the principal 
features of both physical and geometrical optical models of surface reflection
could be accommodated in a so-called unified model which is applicable over 
a wide range of surface roughness and wavelengths.

The unified optical model in {\sc geant4} allows adjustment of parameters to control
the relative contributions from: specular reflections about both the average 
surface normal and the normal of a microfacet at the surface; the diffuse or
Lambertian reflection; a possible backscatter constant; overall 
surface reflectivity.  Figure~\ref{fig:light} shows the polar angular distribution
of light bouncing off such a unified surface, with contributions from the various
sources identified.  Tuning of this model to experimental data~\cite{tyvek2} is 
underway.

\section{Benchmarking}
On a 450~MHz Pentium, our program requires roughly 9 seconds to simulate the tank 
response to a 1~GeV vertical muon\footnote{No optimization was used in the compilation}.  
We anticipate substantial improvement in this figure, possibly through 
application of {\sc top-c}~\cite{topc} or parameterization tools built into the {\sc geant4} framework.

\section{Summary and Prospects}
A prototype simulated water Cerenkov detector has been developed using {\sc geant4}.
We are optimistic about the utility of this package for a number of reasons.

\begin{itemize}
\item The general, extensive arsenal of tools provided will allow 
the detail inherent in the simulation to evolve as needs are identified.
On the other hand, fast parameterizations can be employed in a straightforward
manner wherever details might be found to be unnecessary. 

\item {\sc geant4} design is consistent with modern software trends.
Though the product is relatively new and potholes in the road 
should be expected, the project enjoys a large development community, 
which bodes well for its longevity and evolution.

\item Several shortcomings of {\sc geant3} have been addressed in the new version.
For example the surface reflection model now available is more amenable to 
our problem than its predecessor.  

\end{itemize}

\section*{Acknowledgements}
We would like to thank Peter Gumplinger for help implementing the unified surface model.
This work was supported by the National Science Foundation and CONICET.

\end{document}